\documentclass[twocolumn, prx, reprint, amsfonts, amsmath, amssymb, superscriptaddress, longbibliography, floatfix,aps]{revtex4-1}
\usepackage{amsmath}
\usepackage{amssymb}
\usepackage{braket}
\usepackage{graphicx}
\usepackage[colorlinks,linkcolor=blue,anchorcolor=red,citecolor=green]{hyperref}
\usepackage{pifont}

\usepackage{array, makecell}

\renewcommand{\arraystretch}{2}

\newcolumntype{L}[1]{>{\raggedright\let\newline\\\arraybackslash\hspace{0pt}}m{#1}}
\newcolumntype{C}[1]{>{\centering\let\newline\\\arraybackslash\hspace{0pt}}m{#1}}
\newcolumntype{R}[1]{>{\raggedleft\let\newline\\\arraybackslash\hspace{0pt}}m{#1}}

\newcommand{\cmark}{\ding{51}}
\newcommand{\xmark}{\ding{55}}
\newcommand{\be}{\begin{equation}}
\newcommand{\ee}{\end{equation}}
\newcommand{\bea}{\begin{eqnarray}}
\newcommand{\eea}{\end{eqnarray}}

\newcommand{\la}{\langle}
\newcommand{\ra}{\rangle}

\renewcommand{\vec}[1]{{\bf #1}}

\begin{document}
\title{Kramers {Dichroism} in PT Symmetric Magnets}

\author{Oles Matsyshyn}
\affiliation{Division of Physics and Applied Physics, School of Physical and Mathematical Sciences, Nanyang Technological University, Singapore 637371,
}
\author{Ying Xiong}
\affiliation{Division of Physics and Applied Physics, School of Physical and Mathematical Sciences, Nanyang Technological University, Singapore 637371,
}
\author{Justin C. W. Song}
\email{justinsong@ntu.edu.sg}
\affiliation{Division of Physics and Applied Physics, School of Physical and Mathematical Sciences, Nanyang Technological University, Singapore 637371,
}
\begin{abstract}
{
The quantum states in doubly degenerate Kramers pairs can act as an internal degree of freedom. Here we uncover a {\it Kramers dichroism} in PT symmetric magnets: interband transitions induced by circularly polarized light irradiation can selectively polarize Kramers partnered states allowing to control the Kramers degree of freedom. In contrast, Kramers pairs optically excited by linearly polarized light remain in a completely mixed state. Strikingly, we find a class of second-order nonlinear responses that directly track the polarization between Kramers partnered states. Such Kramers nonlinearities can be pronounced producing large second-order nonlinear layer polarization responses activated by Kramers degeneracy in layered antiferromagnets. Together with Kramers dichroism, these render optical responses a novel means for accessing the Kramers degree of freedom and diagnosing their quantum state. }
\end{abstract}

\maketitle

\section{Introduction}
Spin and its transformation properties under time reversal (T) place strong constraints on electronic states. A case in point are Kramers degeneracies: in the presence of T-symmetry, states with odd half-integer spin occur in orthogonal pairs with the same energy~\cite{RevModPhys.82.3045}. Interestingly, Kramers pairs can persist even in magnetic materials. When both T and inversion, P, symmetries are broken but their composite PT-symmetry is preserved, Kramers degeneracies survive with each electronic Bloch state at least doubly degenerate \cite{Sakurai2020,Tang2016}. Such Kramers pairs are PT-protected~\cite{Gao2021}. As a result, quantum pseudospin polarization between the PT-partnered states can act as an internal Kramers degree of freedom. Nevertheless, in its ground state, PT symmetry ensures each Kramers pair exists in a completely mixed state with zero coherence.

\begin{figure}[t]
    \centering
    \includegraphics[width=0.47\textwidth]{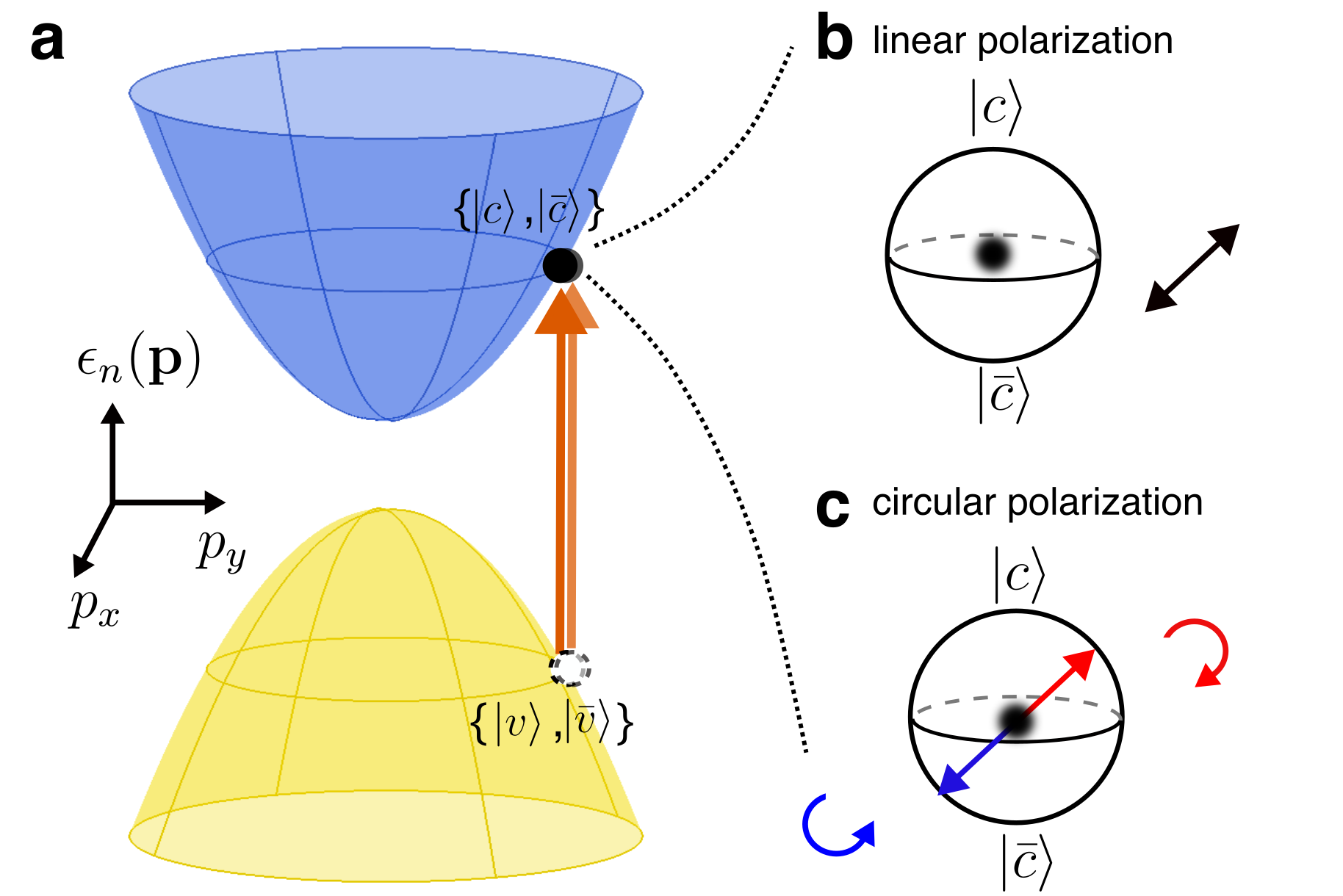}
    \caption{{\bf Kramers Pair Optical Transitions.} 
    {{\bf a.} Optical interband transitions in PT electronic bands involve Kramers {\it pairs} of degenerate states from ($v, \bar{v}$) to ($c, \bar{c}$). {\bf b,c} The quantum state after interband transition can be visualized in a Bloch sphere of the Kramers subspace e.g., for ($c, \bar{c}$). {\bf b.} Linearly polarized light produces a completely mixed state in the Kramers subspace; black blurred dot denotes the trace of the density matrix in the Kramers subspace, Eq.~(\ref{eq:kramersDOF}) {\bf c.} In contrast, circularly polarized light produces a pseudospin polarization of the Kramers partnered states characterized by a Kramers Bloch vector $\boldsymbol{\kappa}$ (red, blue), Eq.~(\ref{eq:kramersDOF}); $\boldsymbol{\kappa}$ can be switched by the helicity of light denoted by red and blue. We term the contrast between the quantum states induced by different light polarizations: ``Kramers Dichroism''.}}
    \label{fig1}
\end{figure}

Here we argue that the Kramers degree of freedom can be directly accessed and addressed by light. We find that when photoexcited by circularly polarized light, a Kramers pseudospin polarization of the PT-partnered Kramers pairs naturally develop, allowing chiral light to control the Kramers degree of freedom. In contrast, linearly polarized light only pumps the total population of the Kramers pair. We term chiral optical control of Kramers pairs ``Kramers dichroism''.

Critically, we identify a class of observables -- PT-odd observables -- that are sensitive to the internal Kramers degree of freedom. PT-even observables such as (the well-studied) charge current~\cite{sodemann2015quantum,matsyshyn2019nonlinear,sipe2000second,de2017quantized,PhysRevX.11.011001,morimoto2016topological,Ahn2022,PhysRevX.10.041041} possess the same expectation value for both states in each Kramers pair and are {\it blind} to the pseudospin polarization between eigenstates in the Kramers pair. In contrast, PT-odd observables (e.g., electric polarization, magnetization, or spin polarization) possess distinct values for states in each Kramers pair allowing to directly probe the Kramers degree of freedom. When combined with Kramers dichroism, we uncover a range of Kramers nonlinearities: second-order nonlinear responses that are sensitive to the Kramers degree of freedom; Kramers nonlinearities are chiral manifesting only for chiral light and switching sign for opposite chiralities. For absorptive events (i.e. resonant transitions) Kramers nonlinearities are the only second-order chiral responses for PT-odd observables. This enables to readily differentiate from other types of responses and renders Kramers nonlinearities a useful means of tracking Kramers dichroism. 

Our work is situated in a current surge of interest in PT symmetric magnetic materials and devices~\cite{Tang2016, Wang2023,Gao2021, Qiu2023,Godinho2018,SuYangXuMBT}. While often treated as a spectator degree of freedom, we find quantum pseudospin polarization sustained between states in the Kramers pairs are essential and can be optically addressed; they lead to pronounced second order nonlinearities that readily manifest in a range of PT-symmetric magnetic materials. For instance, we find even layered MnBi$_2$Te$_4$~\cite{Wang2023,Gao2021} antiferromagnets can realize pronounced Kramers nonlinearity such as a light-induced nonlinear layer polarization that can be switched by both helicity as well as the underlying N\'eel order. This illustrates how Kramers nonlinearities are a powerful means of controlling the layer degree of freedom and probing the underlying antiferromagentic order in quantum materials. 

\section{Kramers degree of freedom and pair transitions}
We begin by analyzing a Bloch Hamiltonian $H_0(\vec p)$ with Bloch eigenstates $|u_n(\vec p)\ra$ and energy $\epsilon_n (\vec p)$. Here $\mathbf{p}$ is the momentum. For materials with PT symmetry, $\mathbf{PT} H_0 (\vec p) (\mathbf{PT})^{-1} = H_0 (\vec p)$ ensures that electrons (odd spin-half particles) occur in doubly-degenerate Kramers pairs: at every $\vec p$, there exist two orthogonal states $|u_n(\vec p)\ra$ and $|u_{\bar n} (\vec p)\ra$ with the same energy $\epsilon_n (\vec p) = \epsilon_{\bar{n}} (\vec p)$. 

This degeneracy produces an internal Kramers degree of freedom. To see this, consider the block density matrix within the Kramers subspace of $ \{|u_n(\vec p)\ra , |u_{\bar n} (\vec p) \ra \}$ that describes its occupation:  
\begin{equation}
\label{eq:kramersDOF}
    \hat{\boldsymbol{\rho}}^{\{n,\bar{n}\}} \hspace{-1mm}= \hspace{-3mm} \sum_{ij \in \{n, \bar n\}} \hspace{-3mm}\rho^n_{ij}| u_i(\vec p)\ra \la u_j(\vec p) |,~  \rho^n_{ij} \hspace{-1mm}=\hspace{-1mm} [\rho_0^n \mathbb{I}+\boldsymbol{\kappa}^n\cdot \boldsymbol{\sigma}^n]_{ij}, 
\end{equation}
$\boldsymbol{\sigma}^{n} \hspace{-1mm}= (\sigma_x, \sigma_y, \sigma_z)$ are Pauli matrices in the $n, \bar{n}$ Kramers subspace, $2\rho_0^n$ is the total occupation of Kramers states, and $\boldsymbol{\kappa}^{n}$ is a three-dimensional Kramers Bloch vector in the Kramers subspace. Since superpositions of Kramers pairs are equally good eigenstates we find $H_0 (\vec p) [\hat{\boldsymbol{\rho}}]_{ij \in \{n, \bar n\}} = \epsilon_n(\mathbf{p}) [\hat{\boldsymbol{\rho}}]_{ij \in \{n, \bar n\}}$. Importantly, the total energy in the Kramers subspace ${\rm Tr} ( H_0 (\vec p)[\hat{\boldsymbol{\rho}}]_{ij \in \{n, \bar n\}} ) = 2\rho^n_0 \epsilon_n(\vec p) $ is {\it blind} to $\boldsymbol{\kappa}^n$. As a result, $\boldsymbol{\kappa}^n$ describes a Kramers degree of freedom: deviations away from equal occupation of the Kramers pair that do not change its energy. 

Notice that $|\boldsymbol{\kappa}^n|$ cannot be gauged away. Even as the Kramers states within the degenerate subspace can be unitarily rotated (e.g. via SU(2) gauge transformation), the Kramers block density matrix contains two gauge invariant quantities: ${\rm Tr}[\hat\rho^n]=2\rho_0^n$ and ${\rm det}[\hat\rho^n]=(\rho_0^n)^2-|\boldsymbol{\kappa}^n|^2$. As a result, $|\boldsymbol{\kappa}^n|$ is also gauge invariant and captures the pseudospin polarization sustained between the Kramers pairs. Notice that finite $|\boldsymbol{\kappa}^n|$ drives Kramers pairs away from a completely mixed state. As a result, $|\boldsymbol{\kappa}^n|$ encodes a type of ``coherence" inside the Kramers pair. Nevertheless, in the ground electronic state of PT-symmetric materials, $\boldsymbol{\kappa}$ is zero for every Kramers pair ensured by its PT symmetry: the ground state is in a {completely} mixed state. 

{Non-zero $\boldsymbol{\kappa}^n$, as we now argue, can be readily induced in excited states by photoexcitation. To show this, we consider the stroboscopic (drive period average) pumping of the Kramers vector $\boldsymbol{\kappa}^n$. Under a periodic drive $\hat V(t) = e \vec E (t) \cdot \hat{\vec r} e^{\eta t}$ with an oscillating electric field $\mathbf{E} = (\mathcal{E}_+e^{-i\omega t}+\mathcal{E}_-e^{i\omega t})/2$ we find the rate of change of the Kramers block density matrix $\langle\partial_t\rho^{n}_{ij}\rangle_t = \int_0^T  [\lim_{\eta\rightarrow0}\partial_t\rho^{n}_{ij}] dt/T$ as 
\begin{align}
\begin{aligned}
\langle\partial_t\rho^{n}_{ij}\rangle_t = \frac{\pi e^2}{2\hbar} \sum_{\nu = \pm 1}\hspace{-1mm}{\mathcal{E}}^\alpha_{-\nu} {\mathcal{E}}^\beta_{\nu}\hspace{-1mm}\sum_{l\in \{l, \bar{l}\}}\hspace{-2mm}r^\alpha_{il}r^{\beta}_{lj}f_{li}\delta(\nu\hbar\omega-\epsilon_{li}),
\end{aligned}
\label{eq:dtkappa}
\end{align}
where $f_{li} (\vec p)= f[\epsilon_l(\vec p)] - f[\epsilon_i(\vec p)]$ is the difference of Fermi functions with $\epsilon_l (\vec p)$ the energy of state with momentum $\vec p$ in band $l \in \{l, \bar{l}\}$. Here the interband Berry connection is $r_{il}^\alpha=A_{il}^\alpha = \la i, \vec p | i\partial^\alpha_p| l, \vec p\ra$ and $\eta$ is an adiabatic turn-on parameter.  }

Eq.~(\ref{eq:dtkappa}) describes Kramers pair transitions. Unlike systems with non-degenerate electronic bands, the interband transitions for Kramers degenerate systems inevitably involve four states: from valence band $\{ |u_v (\vec p)\ra, |u_{\bar v} (\vec p)\ra \}$ subspace to conduction band $\{ |u_c (\vec p)\ra, |u_{\bar c} (\vec p)\ra \}$ subspace, e.g., Fig.~\ref{fig1}a and Eq.~(\ref{eq:dtkappa}) taking $i,j \in \{c,\bar{c}\}$ and $l \in \{ v, \bar{v}\}$. The optical transition process in Kramers degenerate bands described in Eq.~(\ref{eq:dtkappa}) cannot be understood as a simple sum of transitions in two independent sets of non-degenerate bands. Instead, Kramers pair optical transitions are readily understood to occur from one degenerate (valence) subspace to another degenerate (conduction) subspace.

These Kramers pair transitions are essential in capturing a non-zero light induced $\boldsymbol{\kappa}$. Comparing Eq.~(\ref{eq:dtkappa}) with Eq.~(\ref{eq:kramersDOF}), we obtain the rate of change of Kramers Bloch vector magnitude as 
\begin{equation}
\hspace{-2.5mm}\left|\langle\partial_t\boldsymbol{\kappa}^n\rangle_t\right| \hspace{-1mm} = \hspace{-1mm}\sqrt{|w^{n }_{n\bar n}| ^2 +\frac{(w^{n }_{\bar n \bar n}\hspace{-1mm}-\hspace{-1mm}w^{n }_{nn})^2}{4}}, \quad w^{n}_{ij} \equiv \langle\partial_t\rho^{n}_{ij}\rangle_t, 
    \label{eq:ratekappa}
\end{equation}
demonstrating how the light-induced Kramers pair transitions induce pseudospin polarization in the Kramers doublet. Critically, $\left|\langle\partial_t\boldsymbol{\kappa}^n\rangle_t\right|$ in Eq.~(\ref{eq:ratekappa}) is non-zero only for circularly polarized irradiation. In contrast, the achiral component of irradiation pumps $\langle\partial_t\rho^n_0\rangle_t$ the total population of the Kramers subspace, see Appendix~\ref{AsecA} and~\ref{AsecB}. This means that $\partial_t \boldsymbol{\kappa}^n = 0$ for linearly polarized light, with $\boldsymbol{\rho}^{\{n, \bar{n}\}}$ in a completely mixed state. This shows that circularly polarized light irradiation allows to engineer the pseudospin polarization of Kramers pairs, optically controlling from maximally mixed (in the ground state) to partially coherent (in the excited state). 

\section{ Kramers dichroism and PT-odd observables}
How do we track the Kramers degree of freedom encoded in $\boldsymbol{\kappa}$? To proceed we first note that PT symmetry places strong constraints on the matrix elements of PT odd/even observables $ \mathbf{PT} \hat O^{(\pm)} (\mathbf{PT})^{-1} = \pm \hat O^{(\pm)}$ (see Appendix~\ref{AsecA}): 
\be
O_{nn}^{(\pm)} (\vec p) = \pm O_{\bar n \bar n}^{(\pm)} (\vec p), \quad O_{n \bar n}^{(\pm)} (\vec p)= \mp O_{ n \bar n}^{(\pm)} (\vec p), 
\label{eq:Omatrix}
\ee 
where $(\pm)$ superscript indicate observables that are PT even (odd) respectively and ${O}_{nm} (\vec p) = \la n, \vec p | \hat{{O}}| m, \vec p\ra$ is the matrix element for an observable $\hat{{O}}$. Using Eq.~(\ref{eq:Omatrix}) we find PT-even observables $\hat{O}^{(+)}$ are blind to $\boldsymbol{\kappa}^n$. For e.g., the charge current or the charge density $\hat{O}^{(+)} = \{ e\hat{\mathbf{v}}, e\}$ are PT-even. Evaluating their expectation value in the Kramers subspace yields ${\rm Tr} ( \hat{O}^{(+)}\hat{\boldsymbol{\rho}}^{\{n, \bar n\}} ) = 2\rho_0 O_{nn}^{(+)} (\vec{p})$.

In contrast, we find PT-odd observables directly depend on $\boldsymbol{\kappa}^n$.  Evaluating the PT-odd observable expectation value in the $\{ n, \bar{n} \}$ Kramers subspace produces
\begin{equation}
   {\rm Tr}[\hat{O}^{(-)}\hat{\boldsymbol{\rho}}^{\{n,\bar{n}\}}]\hspace{-0.75mm} =\hspace{-0.75mm}  2[\boldsymbol{O}^{(-)}_{n}] \cdot \boldsymbol{\kappa}^n, \quad \hspace{-1mm} \boldsymbol{O}^{(-)}_{n}\hspace{-0.75mm}  =\hspace{-0.75mm} \frac{1}{2} {\rm Tr} [\boldsymbol{\sigma}_n \hat{O}^{(-)}],
   \label{eq:Ominuskappa}
\end{equation}
which depends not only on the magnitude of $\boldsymbol{\kappa}^n$ but also the angle it subtends with $\boldsymbol{O}^{(-)}_{n}$. Notice that such expectations values are impossible without the Kramers subspace; without Kramers degeneracy, matrix elements of $\hat{O}^{-}$ vanish for PT symmetric systems, see Eq.~(\ref{eq:Omatrix}). 

As a result, optically pumped $\boldsymbol{\kappa}^n$ between the Kramers pairs in Eq.~(\ref{eq:ratekappa}) can be directly probed by an injection nonlinearity for PT-odd observables. Using Eq.~(\ref{eq:dtkappa}) and summing over initial and final states we write the injection rate for $O^{(-)}$ observables as 
\begin{align}
\begin{aligned}
\frac{dO^{(-)}}{dt}  \hspace{-0.5mm}=  \hspace{-0.5mm} \frac{\pi e^2}{2\hbar} \hspace{-1mm}\sum_{\nu = \pm 1}\hspace{-1mm}{\mathcal{E}}^\alpha_{-\nu} {\mathcal{E}}^\beta_{\nu}\hspace{-2mm}\sum_{l\in \{l, \bar{l}\}}\hspace{-2mm}r^\alpha_{il}r^{\beta}_{lj}O^{(-)}_{ji}\hspace{-1mm}f_{li}\delta(\nu\hbar\omega\hspace{-0.75mm}-\hspace{-0.75mm}\epsilon_{li}),
\end{aligned}
\label{eq:simpleKramers}
\end{align}
which captures coherent Kramers pair transitions. Eq.~(\ref{eq:simpleKramers}) vanishes for linearly polarized or achiral light but is turned on circularly polarized light directly mirroring the behavior of Eq.~(\ref{eq:ratekappa}). We term the injection rate in Eq.~(\ref{eq:simpleKramers}) ``Kramers injection'' because not only does it depend on $\boldsymbol{\kappa}$, it also {\it vanishes} in PT symmetric systems without Kramers degeneracy. This further underscores the critical role of the Kramers degree of freedom. 

Interestingly, we find that Eq.~(\ref{eq:simpleKramers}) is controlled by $[\mathbf{E}_{-\omega}\times\mathbf{E}_{\omega}]$, see Appendix \ref{AsecB}, changing signs for light with opposite circular polarization. Comparing with Eq.~(\ref{eq:Ominuskappa}), this shows how changing the chirality of circularly polarized light flips the sign of light-induced $\boldsymbol{\kappa}$ and tracks a {\it Kramers dichroism}: opposite chiralities of light produce opposite $\boldsymbol{\kappa}$ vectors. Notice that PT-odd observables are zero in the ground state of PT-symmetric materials. This makes PT-odd observables (e.g., polarization and spin magnetization) ideal markers to track the internal Kramers degree of freedom and its hidden circular (Kramers) dichroism.  

\section{Kramers nonlinearities and quantum geometry}
In the previous section we focused on individual Kramers subspaces to illustrate the role of the Kramers dichroism. The responses associated with Kramers dichroism also naturally emerge from an analysis of the full density matrix where a natural separation into degenerate subspace responses and non-degenerate subspace response emerges; this analysis is discussed in full in Appendix \ref{SecC}. 

We now focus on how PT-odd observables can be connected to quantum geometry \cite{Ahn2022,Provost1980,PhysRevX.11.011001,jankowski2024,ma2022photocurrent,PhysRevLett.131.240001,ScienceAdvancesWang,PhysRevResearch.2.033100, de2017quantized,Rees2020, PhysRevX.10.041041, morimoto2016topological}. We do this by considering the set of Hamiltonians: $H_{\rm eff}(\lambda,\mathbf{p})= H_0(\mathbf{p}) + \lambda \hat O^{(-)}$ parameterized by $\lambda$ with Bloch eigenenergies  $\varepsilon_n(\mathbf{p},\lambda)$ and eigenstates $\ket{ \mathfrak{U}_n(\mathbf{p},\lambda)}$. Writing $\hat O^{(-)} = \partial_\lambda H_{\rm eff}$ produces an identity for $\tilde O_{nm}^{(-)} (\vec p, \lambda) = \bra{\mathfrak{U}_n(\mathbf{p},\lambda)} \hat{O}^{(-)}\ket{ \mathfrak{U}_m(\mathbf{p},\lambda)}$:
\be
\tilde O^{(-)}_{nm} (\vec p, \lambda)= \partial_\lambda \varepsilon_n(\mathbf{p},\lambda)\delta_{nm} + i \varepsilon_{nm}(\mathbf{p},\lambda) \mathcal{A}^\lambda_{nm}(\mathbf{p},\lambda), 
\label{eq:identity}
\ee
where $\varepsilon_{nm} (\mathbf{p},\lambda) = \varepsilon_n (\mathbf{p},\lambda) -\varepsilon_m (\mathbf{p},\lambda)$, and the $\lambda$-Berry connection is $\mathcal{A}^\lambda_{nm} =\bra{\mathfrak{U}_n(\mathbf{p},\lambda)} i\partial_\lambda\ket{ \mathfrak{U}_m(\mathbf{p},\lambda)}$. While $\lambda \neq 0$ breaks PT symmetry thereby lifting degeneracy, $H_{\rm eff}(\lambda \to 0,\mathbf{p}) = H_0(\mathbf{p})$ restores its Kramers pairs. In what follows, we will take $\lambda \to 0$ in our final expressions so that $\tilde O^{(-)}_{nm} (\vec p, \lambda \to 0) = O^{(-)}_{nm} (\vec p)$, with energies $\varepsilon_n (\mathbf{p},\lambda \to 0) = \epsilon_n  (\mathbf{p})$ and Bloch states $\ket{ \mathfrak{U}_m(\mathbf{p},\lambda \to 0)} = \ket{u_m(\mathbf{p})}$. Importantly, geometrical phases accrued in $\mathcal{A}^\lambda_{nm}$ persist even as $\lambda \to 0$, enabling to characterize $O^{(-)}$ nonlinearities with the quantum geometry of its states.

We proceed by systematically enumerating the second-order nonlinearites of PT-odd observables $O^{(-)}$. To do so we employ the standard quantum Liouville treatment of nonlinear response \cite{sipe2000second,matsyshyn2019nonlinear,matsyshyn2023fermi} but generalized to PT symmetric degenerate bands, see Appendix \ref{SecC} for full details. Here we report the main results for Kramers induced second-order nonlinear susceptibilities of $\la O^{(-)} \ra =  e^2\int d\omega~ E^\alpha_{-\omega}E^\beta_{\omega}\chi^{\alpha\beta}$.

\begin{table}[t]
\centering
\resizebox{0.49\textwidth}{!}{
\begin{tabular}{|L{2.8cm}||C{1.25cm}|C{1.25cm}|C{2.cm} | C{2cm}| }
\hline\hline
PT&deg&non-deg&light & equation\\
\hline\hline
$\mathrm{Kramers~Injection}$& \cmark &\xmark& \text{circular} &\text{Eq.(\ref{eq:genOing})}\\\hline
$\mathrm{Kramers~Fermi~Sea}$& \cmark &\xmark& \text{circular} &\text{Eq.(\ref{eq:FS2})}\\\hline
$\mathrm{Shift~Resonant}$& \cmark &\cmark& \text{linear}&\text{Eq.~(\ref{shiftSI})}\\\hline
$\mathrm{Shift~Fermi~Sea}$& \cmark &\cmark& \text{circular}&\text{Eq.~(\ref{shiftFSSI})}\\\hline
\end{tabular}
}
\caption{\label{tab:Table1} Table of $O^{(-)}$ second-order nonlinearities in {PT symmetric systems} that contrast responses of Kramers degenerate (spinful, deg) vs non-degenerate (spinless, non-deg) PT symmetric systems. Here circular indicates difference between left and right hand. Note that while the Shift resonant and Shift Fermi sea responses do not require degeneracy to manifest, nevertheless, their formulae are affected by degeneracy, see Appendices \ref{SecC} and \ref{AsecD}.} 
\end{table}

\begin{figure*}
\centering
\includegraphics[width=0.98\textwidth]{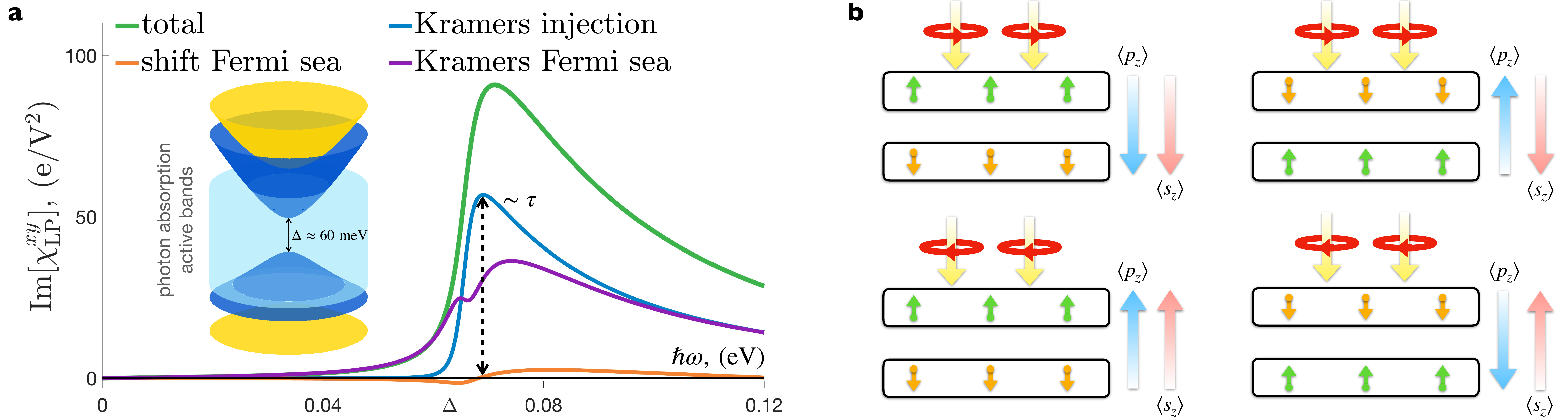}
\caption{{\bf Kramers activated second order nonlinear responses in MBT}. {\bf(a)} Helicity dependent non-linear interlayer polarization susceptibility tensor of bilayer  $\rm Mn Bi_2 Te_4$ ($\tau = 0.25$ps \cite{ma2022photocurrent}, $\mu =0.01$eV, for the complete set of parameters see Appendix \ref{AsecE}). {Note that the KI response is the only circular response in the absorptive regime. While Fermi sea shift responses also display circular polarization sensitivity (orange), their effect is small as compared with KI (blue-cyan) and KFS (purple).} Bilayer MBT band structure (left inset). {\bf{(b)}} Schematic illustration of the interlayer polarization and spin responses for distinct circular (LH/RH) polarizations (red circular arrows) and Néel order parameters $x = \pm 1$, alternating according to the boxed arrow direction.}
\label{FIG2}
\end{figure*}

We identify four second-order nonlinearities of PT-odd observables $\hat O^{(-)}$ (see Table 1) associated with interband processes, see also below. The first is the Kramers injection (KI) nonlinearity [discussed above in Eq.~(\ref{eq:simpleKramers})]. Using Eq.~(\ref{eq:identity}), the Kramers injection nonlinearity can be expressed in a compact geometrical form 
\begin{equation}\label{eq:genOing}
    \chi^{\alpha\beta}_{\rm KI}\hspace{-1mm}=\hspace{-1mm}\frac{\tau\pi}{4}\hspace{-1mm}\int_\mathbf{p}\sum_{nm} \delta(\hbar\omega\hspace{-1mm}-\hspace{-1mm}\varepsilon_{mn})f_{nm}i \hspace{-0.5mm}\left( v^{\alpha;\lambda}_{nm} \mathcal{A}^\beta_{mn}\hspace{-1mm}-\hspace{-0.75mm} v^{\alpha}_{nm} \mathcal{A}^{\beta;\lambda}_{mn}\right),
\end{equation}
where $v^\alpha_{nm}$ is an interband velocity matrix element, and $(;\lambda)$ denotes the generalized derivative 
\begin{equation}
L^{\alpha;\lambda}_{mn} = {\rm lim}_{\lambda \to 0} [D_\lambda,\hat{L}^{\alpha}]_{mn}, \quad D_\lambda = \partial_\lambda-i\mathcal{A}^\lambda \delta^E, 
\end{equation}
of an operator $\hat{L}^\alpha$ (e.g., position or velocity). Note, that ${\rm lim}_{\lambda \to 0} \bra{\mathfrak{U}_n(\mathbf{p},\lambda)} \mathcal{A}^\lambda \delta^E\ket{ \mathfrak{U}_m(\mathbf{p},\lambda)}\equiv \mathcal{A}^\lambda_{nm}\delta_{\epsilon_n\epsilon_m}$ and $[\partial_\lambda,\hat{L}]_{mn} \equiv \partial_\lambda {L}_{mn}$. Importantly, the generalised derivative naturally sums across the degenerate Kramers states: it is U(2) covariant. As a result, when contracted with $v_{nm}^\alpha, \mathcal{A}_{nm}^\beta$, it produces a gauge invariant value. Just like other injection nonlinearities, the KI susceptibility in Eq.~(\ref{eq:genOing}) is extrinsic and depends on an effective relaxation rate $\tau$ or pulse-widths in an ultra-fast measurement \cite{Rees2020,PhysRevResearch.2.033100,PhysRevResearch.2.012017,gao2021intrinsic}. Notice that Kramers degeneracy modifies the constraint $\delta_{\epsilon_n\epsilon_m}$, yielding a shift response consistent with Kramers gauge invariance, see also below.

The second $O^{(-)}$ Kramers nonlinearity we find is a Kramers Fermi sea (KFS) susceptibility that is present even for a filled Fermi sea: 
\begin{equation}\label{eq:FS2}
   \chi^{\alpha\beta}_{\rm KFS}=\frac{\hbar}{2}\int_\mathbf{p}\sum_{nm}f_{nm}i\mathcal{P}\frac{v^{\alpha;\lambda}_{nm} \mathcal{A}^\beta_{mn}- v^{\alpha}_{nm} \mathcal{A}^{\beta;\lambda}_{mn}}{(\hbar\omega-\varepsilon_{mn})^2}. 
\end{equation}
Like the Kramers injection nonlinearity, it similarly vanishes in the absence of Kramers degeneracy for PT symmetric systems. To see this, note that for non-degenerate PT symmetric materials the quantity $v^{\alpha;\lambda}_{nm} \mathcal{A}^\beta_{mn}- v^{\alpha}_{nm} \mathcal{A}^{\beta;\lambda}_{mn}$ when contracted with electric fields becomes proportional to $iA^\alpha_{nm}A^\beta_{mn} (O_{nn}^{(-)} - O^{(-)}_{mm})$; since $O_{nn}^{(-)}$ is zero in PT  symmetric non-degenerate electronic bands, $\chi^{\alpha\beta}_{\rm KFS}$ vanishes. In the same fashion as the KI nonlinearity, KFS only arises for circularly polarized light since $\chi^{\alpha\beta}_{\rm KI, KFS}(\omega)=-\chi^{\alpha\beta}_{\rm KI, KFS}(-\omega)$ in PT symmetric materials: this makes $\chi^{\alpha\beta}_{\rm KI,KFS}$ pure imaginary and antisymmetric under $\alpha \leftrightarrow \beta$. In contrast to KI nonlinearity in Eq.~(\ref{eq:genOing}) above, however, KFS does not require optical absorption and is non-dissipative in nature mirroring a dissipationless photocurrent nonlinearities that have attracted intense recent interest~\cite{matsyshyn2023fermi,shi2023berry,matsyshyn2019nonlinear,PhysRevLett.130.076901}. 

The last two $O^{(-)}$ nonlinearities we find in PT symmetric materials are shift resonant (requires optical absorption) and shift Fermi sea nonlinearities (occurs even for frequencies under the gap), see detailed discussion in Appendix \ref{SecC}. Unlike KI and KFS, such shift resonant and shift Fermi sea $O^{(-)}$ nonlinearities persist even in non-degenerate PT symmetric materials and do not directly track $\boldsymbol{\kappa}$. As shown in Table 1, shift resonant $O^{(-)}$ in PT symmetric systems manifests for linearly polarized light and can be readily distinguished from the KI and KFS responses via helicity dependent irradiation. In contrast, shift Fermi sea $O^{(-)}$ is helicity sensitive. Even as such shift-like nonlinearities do not need Kramers degeneracy to manifest, the presence of Kramers degeneracy nevertheless necessitates modifications to the expressions for shift-like nonlinearities from its naïve non-degenerate form in much the same way as that of gyration photocurrents~\cite{PhysRevX.11.011001}. For a discussion of how the shift formulae are affected by degeneracy, see Appendices \ref{SecC} and \ref{AsecD}.

We note that the interband nonlinearities discussed above are present for both insulators as well as metals; in metals, however, additional nonlinearities associated with intraband processes become activated \cite{PhysRevB.107.205306}.

Lastly, we note that while we have focused on presenting our results for $\hat{O}^{(-)}$ nonlinearities in the main text, our general density matrix formulation that treats both non-degenerate and degenerate cases found in Appendix \ref{SecC} can be used for other observables as well. See Appendices \ref{SecC} and \ref{AsecD} for a full discussion including a discussion on its correspondence with the literature \cite{PhysRevX.10.041041,PhysRevX.11.011001}.

\section{Light induced nonlinear polarization in MnBi$_2$Te$_4$}
To illustrate Kramers {dichroism}, we examine even-layer MnBi$_2$Te$_4$ (MBT): a PT symmetric antiferromagnet. For simplicity, we focus on bilayer MBT; our qualitative conclusions are valid for other even layers. The electronic states of bilayer MBT can be captured by an effective 8-band Hamiltonian \cite{PhysRevLett.124.126402}: $H_0^{\rm MBT} (\mathbf{p}) = [h_1(\mathbf{p}),~t ~; t^\dagger,~h_2(\mathbf{p})]$. Here $h_{l}(\mathbf{p})$ are layer-Hamiltonians (layer index $l\in \{1,2\}$) that can be written as $h_l = h_N(\mathbf{p}) -x(-1)^l h_{\rm AFM}(\mathbf{p})$ \cite{SuYangXuMBT,PhysRevLett.124.126402}; $t$ describes a $\vec{p}$-independent inter-layer coupling that is PT symmetric, see Appendix \ref{AsecE}.

$h_N (\vec p)$ describes a normal component that is the same in both layers: it is P and T invariant. It takes on the familiar BHZ form \cite{opologicalInsulatorsSuperconductorsAndreiBernevig}: $h_N(\mathbf{p}) = \epsilon_0(\mathbf{p})\tau_0\sigma_0+m(\mathbf{p})\tau_z\sigma_z+ v(p_x\tau_0\sigma_y+p_y\tau_0\sigma_x)$ with $\epsilon_0(\vec p)$ and $m(\vec p)$ the dispersive and massive contributions, and $v$ is a velocity. Here $\boldsymbol{\tau}$ and $\boldsymbol{\sigma}$ are Pauli matrices that describe orbital and spin degree of freedom in each layer. In contrast, $h_{\rm AFM}=\frac12{\rm diag}[m_1,-m_2,m_2,-m_1]$ captures the N\'eel order; $x=\pm 1$ captures two distinct antiferromagnetic groundstates with the same energy (see Fig.~\ref{FIG2}). Indeed, $-x(-1)^lh_{\rm AFM}$ breaks P and T symmetry but preserves PT symmetry. $H_0^{\rm MBT} (\mathbf{p})$ produces a bandstructure that is doubly-degenerate at each $\vec p$ (Fig.~\ref{FIG2} inset). Material parameters can be found in Appendix \ref{AsecE}. 

To demonstrate Kramers dichroism in MBT, we now focus on circularly polarized light induced $\hat O^{(-)}$ nonlinearities. These nonlinearities are PT-odd. This is distinct from $\hat{O}^{(+)}$ observables such as the current \cite{PhysRevX.11.011001} or indeed the antiferromagentic N\'eel order both of which are PT-even. We note that helicity-dependent optical control of antiferromagnetic order in MBT has recently been demonstrated experimentally~\cite{Qiu2023}.

A natural $\hat O^{(-)}$ observable in MBT is its layer polarization (LP)~\cite{PhysRevB.107.205306,PhysRevLett.124.077401}, $\hat p_z= {\rm diag}[\mathbb{I}_4,-\mathbb{I}_4]$. LP is odd under PT and possesses an expectation value that vanishes in the equilibrium state of PT symmetric MBT. We numerically compute the $O^{(-)}$ nonlinearities described above of LP for $H_0^{\rm MBT}$ in Fig.~\ref{FIG2}a fixing the Fermi energy in the gap, see Appendix \ref{AsecE} for numerical details. We find KI (blue)/KFS (purple) only manifest as a helicity dependent response described by ${\rm Im}[\chi^{xy}_{\rm LP}(\omega)]$, and display a large and peaked response for frequencies close to transitions about the band extrema. In contrast, the {shift Fermi sea} nonlinear response for LP (shown in red), are almost two orders of magnitude smaller demonstrating the dominant role that KI/KFS play in $\hat O^{(-)}$ nonlinearities allowing track the Kramers dichroism. {Naturally $\tau$ plays a critical role in the size of the nonlinearities. Here we take on a phenomenological approach: in the Appendix \ref{AsecF} we plot other $\tau$ values from $50\, {\rm fs}$ to $1\, {\rm ps}$ that indicate that KI/KFS readily dominate over the shift Fermi sea response \cite{ma2022photocurrent}.}  

Another natural $\hat O^{(-)}$ observable in MBT is spin polarization (SP): $\hat s_z= \frac\hbar2\cdot{\rm diag}[\sigma_z \tau_0,\sigma_z \tau_0]$. Similar to LP, Kramers SP nonlinearities are also helicity dependent, Fig.~\ref{FIG2}b. However, unlike SP, ${\rm Im}[\chi^{xy}_{\rm LP}]$ exhibits a N\'eel order sensitivity, flipping sign when N\'eel order changes from $ x= +1$ to $x =  -1$ (Fig.~\ref{FIG2}{\color{blue}b}). In contrast, ${\rm Im}[\chi^{xy}_{\rm SP}]$ retains its sign for both N\'eel orders; instead its sign is locked to the helicity of the incident EM field. This distinction between $\hat O^{(-)}$ observables comports with the transformation properties of $\hat p_z$ and $\hat s_z$: P flips $\hat p_z$ and N\'eel order but preserves light helicity, while T flips $\hat s_z$, N\'eel order, as well as light helicity. 

{Kramers dichroism affords the ability to optically induce pseudospin polarization of Kramers partnered states in PT symmetric magnets. While we have focused on the above bandgap noninteracting single-particle interband transitions, we anticipate pseudospin polarization between Kramers partners may persist also in the presence of interactions: e.g., in excitonic bound states which are formed from the superposition of particle-hole excitations between valence and conduction bands.} Importantly, intra-Kramers-pair pseudospin polarization produce of a class of second order nonlinearities that allow to directly track optically induced superpositions. We find they dominate the nonlinear response of layer polarization responses in MBT antiferromagnets. For instance, Kramers polarization susceptibilities ${\rm Im}[\chi^{xy}_{\rm LP}]$ of order $100\, {\rm e}/{\rm V}^2$ that readily dominate the interlayer polarization response of MBT; such KI/KFS nonlinearities are orders of magnitude larger than those recently found for twisted bilayer graphene~\cite{PhysRevLett.124.077401}. Analogous to other optical responses, KI/KFS nonlinearities may also experience enhancements close to phase transitions \cite{PhysRevLett.116.237402} that can be readily controlled in 2D magnetic platforms \cite{doi:10.1126/sciadv.aaw5685}.  Kramers dichroism and nonlinearity together form a toolbox that can be used to prepare and map out superpositions of Kramers quantum states. We anticipate that this can open up the possibility of exploiting such coherences in a new type of quantum optoelectronics based off the Kramers degree of freedom.

\begin{acknowledgments}
This research was supported by the Singapore Ministry of Education (MOE) Academic Research Fund Tier 2 grant (MOE-T2EP50222-0011) and Tier 3 grant (MOE-MOET32023-0003) “Quantum Geometric Advantage”.
\end{acknowledgments}
\appendix

\section{Symmetry properties of operators and zero Kramers Bloch vector in PT symmetric ground state}\label{AsecA}

In this appendix we will examine the constraints that PT symmetry places on states and matrix elements. We first note that for PT partnered states $|u_n\rangle$ and $|u_{\bar{n}}\rangle$, we can write 
\begin{equation}
    PT|u_n\rangle = e^{i\phi_{n \bar{n}}} |u_{\bar{n}}\rangle, \quad PT|u_{\bar{n}}\rangle = e^{i\phi_{\bar{n} n}} |u_n\rangle, 
\end{equation}
where $\phi$ angles are a general phase factor associated with the transformation. Since $(PT)^2 = -1$ (for fermionic electronic states), we have $(PT)^2|u_n\rangle = PTe^{i\phi_{n \bar{n}}} |u_{\bar{n}}\rangle= e^{-i\phi_{n \bar{n}}} PT|u_{\bar{n}}\rangle$ thus $e^{i( \phi_{\bar{n} n}-\phi_{n \bar{n}})} = -1.$ 

Using the above we can now examine the constraints that PT symmetry imposes on matrix elements of an operator $\hat{O}$. We will consider two sets of operators:  PT-even ($\xi = +1$) and PT-odd ($\xi=-1$) so that $PT\hat{O}^{(\xi)}(PT)^{-1} = \xi \hat{O}^{(\xi)}$. We find matrix elements of this operator satisfy: 
\begin{equation}\label{PTtrans}
\begin{aligned}
    \langle u_m | \hat{O} | u_n \rangle &= e^{i\phi_{\bar{m}{m}}} e^{-i\phi_{\bar{n}{n}}} \xi \langle PT u_{\bar{m}} | PT \hat{O} (PT)^{-1} | PT u_{\bar{n}} \rangle \\&= e^{i(\phi_{\bar{m}{m}}-\phi_{\bar{n}{n}})}  \xi \langle PT u_{\bar{m}} | PT \hat{O} | u_{\bar{n}} \rangle \\&= e^{i(\phi_{\bar{m}{m}}-\phi_{\bar{n}{n}})} \xi \langle \hat{O} u_{\bar{n}} | u_{\bar{m}} \rangle \\&= e^{i(\phi_{\bar{m}{m}}-\phi_{\bar{n}{n}})} \xi \langle u_{\bar m} | \hat{O} | u_{\bar n} \rangle^*\\&=e^{i(\phi_{\bar{m}{m}}-\phi_{\bar{n}{n}})}\xi\langle u_{\bar n} | \hat{O} | u_{\bar m} \rangle.
\end{aligned}
\end{equation}
Using Eq.~(\ref{PTtrans}) we can directly examine matrix elements in the Kramers subspace $\{ n, \bar{n}\}$. Substituting the diagonal case $m=n$ as well as off-diagonals $m=\bar{n}$ we obtain Eq.~(\ref{eq:Omatrix}), of the main text reproduced here for the convenience of the reader: 
\be
O_{nn}^{(\pm)} (\vec p) = \pm O_{\bar n \bar n}^{(\pm)} (\vec p), \quad O_{n \bar n}^{(\pm)} (\vec p)= \mp O_{ n \bar n}^{(\pm)} (\vec p), 
\ee 
Importantly, we find that off-diagonal matrix elements for PT-even operators $O_{n \bar n}^{(+)}$ vanish. Similarly, its on-diagonal matrix elements are equal. 

This can be directly applied to the density matrix in a PT symmetric ground state. Notice that in the PT symmetric ground state, the density matrix $\mathbf{PT}\hat{\boldsymbol{\rho}}^{\{n,\bar{n}\}}[\mathbf{PT}]^{-1}=\hat{\boldsymbol{\rho}}^{\{n,\bar{n}\}}$ is PT even. Hence, its off-diagonal matrix elements must vanish. As a result, $\boldsymbol{\kappa}^{n}_{x,y}$ vanish. Similarly, its PT-even character means that its on-diagonal matrix elements in the Kramers subspace are equal. As a result, we conclude that $\boldsymbol{\kappa}^{n}_{z} =0$ also vanishes in the PT symmetric ground state. Notice $\boldsymbol{\kappa}$ can be non-vanishing in the excited state. Indeed, as discussed in the main text, Kramers dichroism allows finite $\boldsymbol{\kappa}$ to develop. 

\section{Rate of Change of Kramers Block Density Matrix}\label{AsecB} 

The density matrix in the Kramers subspace can be directly obtained using a standard quantum Liouville approach (see discussion below). Focusing on the Kramers subspace $\{n, \bar{n}\}$ (i.e. $\epsilon_m(\mathbf{p}) = \epsilon_{{n}}(\mathbf{p})$) we find:
\begin{multline}
    \rho^{(2)}_{nm \in \{ n, \bar{n}\}}(t)\hspace{-0.75mm}=\hspace{-1.25mm}\iint_{\omega_1,\omega_2}\hspace{-6.25mm}e^{-i(\omega_1+\omega_2)t+2\eta t/\hbar}\frac{e^2 E^{\alpha}(\omega_2)E^{\beta}(\omega_1)}{\hbar(\omega_1+\omega_2)+2i\eta} \\\times\sum_c\bigg[\frac{r^\beta_{nc}r^{\alpha}_{cm}f_{nc}}{\hbar\omega_2-\epsilon_{cn}+i\eta}-\frac{r^\alpha_{nc}r^{\beta}_{cm}f_{cn}}{\hbar\omega_2-\epsilon_{nc}+i\eta}\bigg],
\end{multline} 
where $\int_\omega \equiv \int_{-\infty}^\infty d\omega/(2\pi)$. For monochromatic light $\mathbf{E}(\omega_i) = 2\pi\sum_{\nu = \pm 1}\boldsymbol{\mathcal{E}}_\nu\delta(\nu\omega-\omega_i)/2$, with $\boldsymbol{\mathcal{E}}_{-} = (\boldsymbol{\mathcal{E}}_+)^*$, the Kramers block density matrix above has the following rate of change:
 \begin{multline}
    \partial_t\rho^{(2)}_{nm\in \{ n, \bar{n}\}}(t)\hspace{-1mm}=\hspace{-1mm} \frac{-ie^2}{4\hbar}\sum_{\nu,\nu' = \pm 1}\hspace{-3mm}{\mathcal{E}}^\alpha_\nu {\mathcal{E}}^\beta_{\nu'}e^{-i(\nu+\nu')\omega t+ 2\eta t/\hbar}\times\\\times \sum_c\bigg[\frac{r^\beta_{nc}r^{\alpha}_{cm}f_{nc}}{\hbar\nu\omega-\epsilon_{cn}+i\eta}-\frac{r^\alpha_{nc}r^{\beta}_{cm}f_{cn}}{\hbar\nu\omega-\epsilon_{nc}+i\eta}\bigg]\label{raterho}
\end{multline} 
Focusing on the DC part (rectified, e.g. $\nu+\nu'=0$) we find Eq.~(\ref{eq:dtkappa}) of the main text. We note that this Kramers block density matrix is $2 \times 2$. To facilitate comparison with Eq.~(\ref{eq:kramersDOF}), we decompose into a trace and Kramers Bloch vector components as
\begin{equation}\label{KRamersBlock}
\langle\partial_t\boldsymbol{\kappa}^n\rangle_t\hspace{-0.05mm}\cdot\hspace{-0.05mm}\boldsymbol{\sigma}\hspace{-1mm}=\hspace{-1mm}\int_0^T\hspace{-2mm}\frac{dt}{T}[\partial_t\boldsymbol{\kappa}^n\hspace{-0.05mm}\cdot\hspace{-0.05mm}\boldsymbol{\sigma}] \hspace{-1mm}=\hspace{-1mm} \left(\hspace{-2mm}\begin{array}{cc}\frac{w^{n}_{nn}-w^{n}_{\bar n \bar n}}{2} & w^{n }_{n\bar n} \\w^{n }_{\bar n n}& \frac{w^{n}_{\bar n \bar n}-w^{n }_{nn}}{2}
    \end{array}\hspace{-1mm}\hspace{-1mm}\right),
\end{equation}
where $2\langle\partial_t\rho^n_0\rangle_t = w^{n}_{\bar n \bar n}+w^{n }_{nn}$ and $w^{n}_{ij} \equiv \langle\partial_t\rho^{n}_{ij}\rangle_t$ denote individual components of the {\it rate of change} of the Kramers block density matrix written out for the convenience of the reader
\begin{align}
\begin{aligned}
w^{n }_{ij}&= \frac{\pi e^2}{2\hbar }\sum_{\nu = \pm 1}{\mathcal{E}}^\alpha_{-\nu} {\mathcal{E}}^\beta_{\nu}\sum_cr^\alpha_{ic}r^{\beta}_{cj}f_{ci}\delta[\epsilon_{ci}\hspace{-0.5mm}-\hspace{-0.5mm}\nu\hbar\omega]\\&= \frac{\pi e^2}{2\hbar\omega^2}\sum_{\nu = \pm 1}{\mathcal{E}}^\alpha_{-\nu} {\mathcal{E}}^\beta_{\nu}\sum_cv^\alpha_{ic}v^{\beta}_{cj}f_{ci}\delta[\epsilon_{ci}\hspace{-0.5mm}-\hspace{-0.5mm}\nu\hbar\omega],
\end{aligned}
\label{eq:kappaSI}
\end{align}
where after relabeling dummy index $(\alpha\leftrightarrow\beta)$ we have also used the identity $\mathbf{v} = (i/\hbar) [\mathbf{r}, H]$ where $\mathbf{v} = \hbar^{-1}\partial_\mathbf{p}\hat H(\mathbf{p})$ and for $\epsilon_i(\mathbf{p})\neq\epsilon_c(\mathbf{p})$ we have $\hbar v^\alpha_{ic} = i \epsilon_{ic} r^\alpha_{ic}$. 

Applying Eq.(\ref{PTtrans}) above we find that non-zero $\boldsymbol{\kappa}^n$ can be accessed only via the circularly polarized light irradiation. For the convenience of the reader, we show this explicitly. First we apply Eq.(\ref{PTtrans}) onto 
\begin{align}\label{useptoff}
\begin{aligned}
&\sum_{\nu = \pm 1}\hspace{-1.5mm}{\mathcal{E}}^\alpha_{-\nu} {\mathcal{E}}^\beta_{\nu}\hspace{-1.mm}\sum_cv^\alpha_{ic}v^{\beta}_{cj}f_{ci}\delta[\epsilon_{ci}\hspace{-0.5mm}-\hspace{-0.5mm}\nu\hbar\omega]=\\&=\sum_{\nu = \pm 1}\hspace{-1.5mm}{\mathcal{E}}^\alpha_{-\nu} {\mathcal{E}}^\beta_{\nu}\hspace{-1.mm}\sum_{c}e^{i(\phi_{\bar i i}-\phi_{\bar j j})}v^\alpha_{\bar c \bar i}v^{\beta}_{\bar j \bar c}f_{ci}\delta[\epsilon_{ci}\hspace{-0.5mm}-\hspace{-0.5mm}\nu\hbar\omega]=\\&=\sum_{\nu = \pm 1}\hspace{-1.5mm}({\mathcal{E}}^\alpha_{-\nu} {\mathcal{E}}^\beta_{\nu})^*\hspace{-1.mm}\sum_{c}e^{i(\phi_{\bar i i}-\phi_{\bar j j})}v^{\alpha}_{\bar j  c}v^\beta_{ c \bar i}f_{ci}\delta[\epsilon_{ci}\hspace{-0.5mm}-\hspace{-0.5mm}\nu\hbar\omega],
\end{aligned}
\end{align}
where in the last line we used dummy relabeling  $(\alpha\leftrightarrow \beta), (\bar c\rightarrow c)$ does not affect factors that depend on energy only. 

We first examine the off-diagonal elements of the Kramers block density matrix $j = \bar i$ in Eq.(\ref{eq:kappaSI}). Applying Eq.(\ref{useptoff}) [note $e^{i(\phi_{\bar i i}-\phi_{i\bar i)}}=-1$] we find: 
\begin{equation}\label{woffd}
\begin{aligned}
    w^{n }_{i\bar i}\hspace{-1mm}&=\hspace{-1mm} \frac{\pi e^2}{2\hbar\omega^2}\sum_{\nu = \pm 1}\hspace{-1mm}{\mathcal{E}}^\alpha_{-\nu} {\mathcal{E}}^\beta_{\nu}\hspace{-1mm}\sum_cv^\alpha_{ic}v^{\beta}_{c\bar i}f_{ci}\delta[\epsilon_{ci}\hspace{-0.5mm}-\hspace{-0.5mm}\nu\hbar\omega]\\&\hspace{-1mm}=\frac{\pi e^2}{4\hbar\omega^2}\sum_{\nu = \pm 1}\hspace{-1mm}{\mathcal{E}}^\alpha_{-\nu} {\mathcal{E}}^\beta_{\nu}\hspace{-1mm}\sum_cv^\alpha_{ic}v^{\beta}_{c\bar i}f_{ci}\delta[\epsilon_{ci}\hspace{-0.5mm}-\hspace{-0.5mm}\nu\hbar\omega]\\&\quad-\frac{\pi e^2}{4\hbar\omega^2}\sum_{\nu = \pm 1}\hspace{-1mm}({\mathcal{E}}^\alpha_{-\nu} {\mathcal{E}}^\beta_{\nu})^*\hspace{-1mm}\sum_cv^\alpha_{ic}v^{\beta}_{c\bar i}f_{ci}\delta[\epsilon_{ci}\hspace{-0.5mm}-\hspace{-0.5mm}\nu\hbar\omega]\\&\hspace{-1mm}=\frac{i\pi e^2}{2\hbar\omega^2}\sum_{\nu = \pm 1}\hspace{-1mm}{\rm Im}[{\mathcal{E}}^\alpha_{-\nu} {\mathcal{E}}^\beta_{\nu}]\sum_cv^\alpha_{ic}v^{\beta}_{c\bar i}f_{ci}\delta[\epsilon_{ci}\hspace{-0.5mm}-\hspace{-0.5mm}\nu\hbar\omega].
\end{aligned}
\end{equation}
We can apply the same treatment to $\kappa_z$ component:   
\begin{equation}\label{wdiff}
\begin{aligned}
    w^{n }_{ii}\hspace{-0.5mm}-\hspace{-0.5mm}w^{n }_{\bar i\bar i}&\hspace{-1mm}=\hspace{-0.5mm} \frac{\pi e^2}{2\hbar\omega^2}\sum_{\nu = \pm 1}\hspace{-1mm}{\mathcal{E}}^\alpha_{-\nu} {\mathcal{E}}^\beta_{\nu}\hspace{-1mm}\sum_cv^\alpha_{ic}v^{\beta}_{c i}f_{ci}\delta[\epsilon_{ci}\hspace{-0.75mm}-\hspace{-0.75mm}\nu\hbar\omega] \\&\quad- \frac{\pi e^2}{2\hbar\omega^2}\sum_{\nu = \pm 1}\hspace{-1mm}{\mathcal{E}}^\alpha_{-\nu} {\mathcal{E}}^\beta_{\nu}\hspace{-1mm}\sum_cv^\alpha_{\bar ic}v^{\beta}_{c\bar i}f_{ci}\delta[\epsilon_{ci}\hspace{-0.75mm}-\hspace{-0.75mm}\nu\hbar\omega]\\&\hspace{-1mm}= \hspace{-0.5mm}\frac{\pi e^2}{2\hbar\omega^2}\sum_{\nu = \pm 1}\hspace{-1mm}{\mathcal{E}}^\alpha_{-\nu} {\mathcal{E}}^\beta_{\nu}\hspace{-1mm}\sum_cv^\alpha_{ic}v^{\beta}_{c i}f_{ci}\delta[\epsilon_{ci}-\nu\hbar\omega] \\&\quad - \frac{\pi e^2}{2\hbar\omega^2}\sum_{\nu = \pm 1}\hspace{-1mm}({\mathcal{E}}^\alpha_{-\nu} {\mathcal{E}}^\beta_{\nu})^*\hspace{-1mm}\sum_cv^\alpha_{ ic}v^{\beta}_{c i}f_{ci}\delta[\epsilon_{ci}\hspace{-0.75mm}-\hspace{-0.75mm}\nu\hbar\omega]\\&\hspace{-1mm}=\hspace{-0.5mm} \frac{i\pi e^2}{\hbar\omega^2}\sum_{\nu = \pm 1}\hspace{-1mm}{\rm Im}[{\mathcal{E}}^\alpha_{-\nu} {\mathcal{E}}^\beta_{\nu}]\sum_cv^\alpha_{ic}v^{\beta}_{c i}f_{ci}\delta[\epsilon_{ci}\hspace{-0.75mm}-\hspace{-0.75mm}\nu\hbar\omega],
\end{aligned}
\end{equation}
where in the second line we also employed Eq.(\ref{useptoff}). From this calculation we conclude that circularly polarized light controls Kramers degree of freedom in PT materials, see Eq.(\ref{KRamersBlock}). Importantly, as discussed in the main text, {Eqs.(\ref{wdiff}, \ref{woffd})} together with Eq.(\ref{KRamersBlock}) demonstrate the direction of the Kramers vector is reversed under switching the circular light polarization. 

In contrast, for rate of change of the trace of the Kramers block density matrix we find:
\begin{equation}
    w^{n }_{ii}+w^{n }_{\bar i\bar i}\hspace{-1mm}=\hspace{-1mm}\frac{\pi e^2}{\hbar\omega^2}\sum_{\nu = \pm 1}{\rm Re}[{\mathcal{E}}^\alpha_{-\nu} {\mathcal{E}}^\beta_{\nu}]\sum_cv^\alpha_{ic}v^{\beta}_{c i}f_{ci}\delta[\epsilon_{ci}-\nu\hbar\omega],
\end{equation}
which is light helicity blind.

\section{Systematic quantum Liouville approach for nonlinear response}\label{SecC}
In order to systematically derive the optical response of an electronic insulator (being careful to include the effects of Kramers degeneracy), we perturbatively solve the Liouville equation $ i\hbar d_t\hat{\rho}(\mathbf{k},t)= [\hat{H}_0(\mathbf{k}) + \hat V(\mathbf{k},t),\hat{\rho}(\mathbf{k},t)],$ assuming that in equilibrium Bloch states are thermally populated $\bra{n(\mathbf{k})}\hat{\rho}_{\rm eq}(\mathbf{k})\ket{m(\mathbf{k})} = f_{\rm FD}[\epsilon_n(\mathbf{k})]\delta_{nm}$. Here $\hat{\rho}(\mathbf{k},t)$ is the density matrix. At second order, the correction to the rectified density matrix in the frequency domain is 
\begin{multline}\label{eq:fullIntDM}
    \hat{\rho}^{(2)}_{mn}=e^2\int_\omega E^{\alpha}_{-\omega}E^{\beta}_{\omega}e^{2\eta t/\hbar}\times\\\times\Bigg[
    \frac{1}{\epsilon_{nm}+2i\eta}i\partial^{\alpha}_k\frac{A^{\beta}_{mn}f_{nm}}{\hbar\omega-\epsilon_{mn}+i\eta}+\\\frac{1}{\epsilon_{nm}+2i\eta}\sum_c\hspace{-1mm}\left\{\frac{A^\alpha_{mc}A^{\beta}_{cn}f_{nc}}{\hbar\omega-\epsilon_{cn}+i\eta}\hspace{-0.75mm}-\hspace{-0.75mm}\frac{A^\beta_{mc}A^{\alpha}_{cn}f_{cm}}{\hbar\omega-\epsilon_{mc}+i\eta}\right\}\hspace{-1mm}\Bigg].
\end{multline}
The nonlinear rectified response of any observable $O$ represented by an operator $\hat O$ is given by $\langle O \rangle = \int_\mathbf{p}\sum_{nm}O_{nm}\rho_{mn}$. We write such explicitly by contracting Eq.(\ref{eq:fullIntDM}) with $O_{nm}$ to find:
\begin{multline}\label{eq:interres}
     O^{(2)}_{\rm inter}= e^2\int_{\vec p}\int_\omega E^\alpha_{-\omega}E^\beta_{\omega}e^{2\eta t/\hbar}\times\\\times\sum_{nm}\Bigg[\frac{O_{nm}}{\epsilon_{nm}+2i\eta}i\partial^\alpha_k\frac{A^\beta_{mn}f_{nm}}{\hbar\omega+\epsilon_{nm}+i\eta}+\\+\frac{A^\beta_{mn}f_{nm}}{\hbar\omega+\epsilon_{nm}+i\eta}{\sum_{c}}\left\{\frac{ O_{nc}A^\alpha_{cm}}{\epsilon_{nc}+2i\eta}-\frac{A^\alpha_{nc} O_{cm}}{\epsilon_{cm}+2i\eta}\right\}\Bigg].
\end{multline}
We note that the nonlinear susceptibility tensor  $\chi^{\alpha\beta}(\omega)$, there is a ``gauge'' freedom. This is because when $\chi^{\alpha\beta}(\omega)$ contracted with a pair of electric fields, the physical response is symmetric under $\alpha\leftrightarrow\beta,\omega\leftrightarrow-\omega$: 
\begin{multline}
    \sum_{\alpha\beta}\int_\omega~ E^\alpha_{-\omega}E^\beta_{\omega}\chi^{\alpha\beta}(\omega) =\\= \frac{1}{2}\sum_{\alpha\beta}\int_\omega~ E^\alpha_{-\omega}E^\beta_{\omega}(\chi^{\alpha\beta}(\omega)+\chi^{\beta\alpha}(-\omega)). 
\end{multline}
In what follows, as well as in the main text, we write the susceptibility tensors in a {\it symmetrized} form.

Interestingly, altermagnets \cite{PhysRevX.12.040501} provide another interesting platform for physics involving degenerate pairs, as XT-protected degeneracies could in principle enable analogous pseudospin polarization-induced nonlinear responses. However, their quantitative contribution depends on the phase-space measure of the degenerate manifold. When degeneracies occur only at isolated points or lines, their contribution is expected to be small.

\subsection{Kramers nonlinearities}
The different terms of Eq.(\ref{eq:interres}) represent distinct nonlinearities. We first focus on the second term in the square brackets of Eq.(\ref{eq:interres}). Importantly, first term in curly brackets when $\epsilon_c=\epsilon_n$ and second term for $\epsilon_c=\epsilon_m$ are proportional to $1/\eta$. For small $\eta$ these terms are large. As we will now see, these terms form Kramers nonlinearities. Focusing on these nonlinear response terms in Eq.(\ref{eq:interres}) and writing them in a \textit{symmetrised} form, we obtain 
\begin{multline}\label{supKR}
     O^{(2)}_{\rm Kramers}= \frac{e^2}{2}\int_{\vec p}\int_\omega E^\alpha_{-\omega}E^\beta_{\omega}e^{2\eta t/\hbar}\times\\\sum_{nm}\frac{A^\beta_{mn}f_{mn}}{(\hbar\omega+\epsilon_{nm})^2+\eta^2}\left\{\sum_{c,\epsilon_c=\epsilon_n}\hspace{-2.25mm}O_{nc}A^\alpha_{cm}\hspace{-0.75mm}-\hspace{-1.75mm}\sum_{c,\epsilon_c=\epsilon_m}\hspace{-2.5mm}A^\alpha_{nc} O_{cm}\right\}.
\end{multline}
Next we observe the following identity: 
\begin{multline}\label{res_fs}
     \lim_{\eta\rightarrow0}\frac{1}{(\hbar\omega+\epsilon_{nm})^2+\eta^2}=\\=\lim_{\eta\rightarrow0}\frac{\eta^2}{((\hbar\omega+\epsilon_{nm})^2+\eta^2)^2}+\lim_{\eta\rightarrow0}\frac{(\omega+\epsilon_{nm})^2}{((\hbar\omega+\epsilon_{nm})^2+\eta^2)^2} =\\= \frac{\pi}{2\eta}\delta(\hbar\omega+\epsilon_{nm})+\mathcal{P}\frac{1}{(\hbar\omega+\epsilon_{nm})^2},
\end{multline}
Applying the above identity to Eq.~(\ref{supKR}) we obtain the Kramers Injection (KI) nonlinearity as
\begin{multline}
    O^{(2)}_{\rm KI}= \frac{\pi e^2}{4\eta}e^{2\eta t/\hbar}\int_\omega E^\alpha_{-\omega}E^\beta_{\omega}\int_{\vec p}\sum_{nm}\delta(\hbar\omega-\epsilon_{mn})\times\\\times f_{mn}A^\beta_{mn}\left\{\sum_{c,\epsilon_c=\epsilon_n}\hspace{-2.5mm}O_{nc}A^\alpha_{cm}-\hspace{-1.25mm}\sum_{c,\epsilon_c=\epsilon_m}\hspace{-2.5mm}A^\alpha_{nc} O_{cm}\right\}, 
\label{eq:FullKI}
\end{multline}
and Kramers Fermi Sea (KFS) nonlinearity as 
\begin{multline}
    O^{(2)}_{\rm KFS} = \frac{e^2}{2}\int_\omega E^\alpha_{-\omega}E^\beta_{\omega}e^{2 \eta t/\hbar}\int_{\vec p}\sum_{nm}\mathcal{P}\frac{A^\beta_{mn}f_{mn}}{(\hbar\omega-\epsilon_{mn})^2}\\\left\{\sum_{c,\epsilon_c=\epsilon_n}O_{nc}A^\alpha_{cm}-\sum_{c,\epsilon_c=\epsilon_m}A^\alpha_{nc} O_{cm}\right\}.
\label{eq:FullKFS}
\end{multline}
Note that taking a time derivative of the Kramers injection nonlinearity in Eq.~(\ref{eq:FullKI}) in the limit of $\eta \to 0$ yields Eq.~(\ref{eq:simpleKramers}) of the main text. This corresponds to the resonant (absorptive) part of of Eq.~(\ref{supKR}). As we will see below, by using the $\lambda$-formulation, the quantum geometric content of Eq.~(\ref{eq:FullKI}) can be revealed see e.g., Eq.~(\ref{eq:genOing}) of the main text. 

Eq.~(\ref{eq:FullKFS}) corresponds to the Fermi sea component of Eq.(\ref{supKR}) and arises even in the absence of any absorption. By using the $\lambda$-formulation, this KFS nonlinearity in Eq.~(\ref{eq:FullKFS}) produces Eq.~(\ref{eq:FS2}) of the main text. The above forms are numerically friendly, and was employed to produce Fig.~2 of the main text. 

\begin{figure*}
\centering
\includegraphics[width=0.9\textwidth]{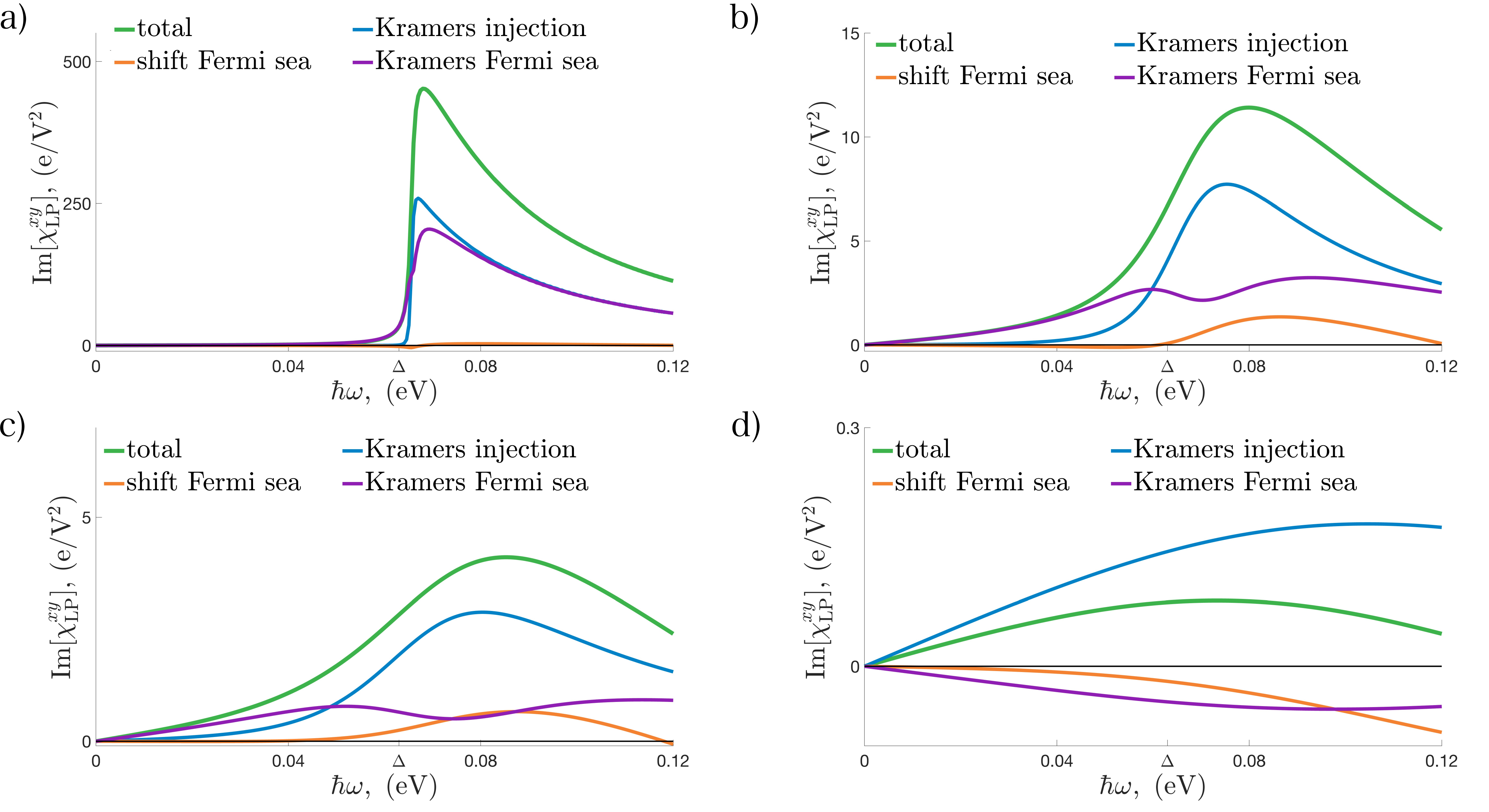}
\caption{{\bf  Second order nonlinear responses for PT-odd observables in MBT}.  Helicity dependent non-linear interlayer polarization susceptibility tensor of bilayer  $\rm Mn Bi_2 Te_4$ for $\mu =0.01$eV and {\bf(a)} $\tau = 1$ ps,  {\bf(b)} $\tau = 50$ fs,  {\bf(c)} $\tau = 25$ fs,  {\bf(d)} $\tau = 1$ fs; for the complete set of parameters see Appendix \ref{AsecE}. Similar to the result displayed on Fig.\ref{FIG2}a of the main text, for the relaxation times shown, Kramers nonlinearities (KI and KFS) are larger than the shift Fermi sea nonlinearity. Notice that shift resonant contribution is helicity blind, as a result $\frac{1}{2}\delta O^{(2)}_{\rm SR} ({\rm LH}) - \frac{1}{2}\delta O^{(2)}_{\rm SR} ({\rm RH})$ vanishes in a PT symmetric material .}
\label{FIGS1}
\end{figure*}

\subsection{Shift resonant and shift Fermi sea nonlinearities}
Considering the rest of the terms in Eq.(\ref{eq:interres}) we find two sets of shift nonlinearities: a shift resonant and a shift Fermi sea contribution. After integrating by parts for $\partial^\alpha_k$ in Eq.(\ref{eq:interres}), we find a shift resonant (SR) and shift Fermi sea (SFS) components as:
\begin{multline}\label{shiftSI}
    \delta O^{(2)}_{\rm SR}= \pi e^2\int_\omega E^\alpha_{-\omega}E^\beta_{\omega}\int_{\vec p}\sum_{nm}\delta(\hbar\omega-\epsilon_{mn})A^\beta_{mn}f_{mn}\times\\\times\Bigg[\partial^\alpha_\mathbf{k}\frac{O_{nm}}{\epsilon_{nm}}-i\left\{\sum_{c,\epsilon_c\neq \epsilon_m}\hspace{-2.25mm}A^\alpha_{nc}\frac{ O_{cm}}{\epsilon_{cn}}-\hspace{-2.25mm}\sum_{c,\epsilon_c\neq \epsilon_n}\frac{ O_{nc}}{\epsilon_{nc}}A^\alpha_{cm}\right\}\Bigg].
\end{multline}
\begin{multline}\label{shiftFSSI}
    \delta O^{(2)}_{\rm SFS}= e^2\int_\omega E^\alpha_{-\omega}E^\beta_{\omega}\int_{\vec p}\sum_{nm}\mathcal{P}\frac{iA^\beta_{mn}f_{mn}}{\hbar\omega-\epsilon_{mn}}\times\\\times\Bigg[\partial^\alpha_\mathbf{k}\frac{O_{nm}}{\epsilon_{nm}}-i\left\{\sum_{c,\epsilon_c\neq \epsilon_m}\hspace{-2.25mm}A^\alpha_{nc}\frac{ O_{cm}}{\epsilon_{cn}}-\hspace{-2.25mm}\sum_{c,\epsilon_c\neq \epsilon_n}\frac{ O_{nc}}{\epsilon_{nc}}A^\alpha_{cm}\right\}\Bigg].
\end{multline}
where we have used the identity: $\lim_{\eta\rightarrow0} {1}/({x+i\eta}) = \mathrm{P}({1}/{x})-i\pi\delta(x)$ to separate the contributions. 

A numerically friendly form of the expression in the brackets is:
\begin{multline}
    \partial^\beta_{\mathbf{k}} \frac{O_{mn}}{\epsilon_{mn}}-i\left[{\sum_{c,\epsilon_c\neq\epsilon_n}}\hspace{-1mm}A^\beta_{mc}\frac{ O_{cn}}{\epsilon_{cn}}-\hspace{-1mm}{\sum_{c,\epsilon_c\neq\epsilon_m}}\frac{ O_{mc}}{\epsilon_{mc}}A^\beta_{cn}\right]=\\=
    \frac{\bra{m}\frac{\partial \hat O}{\partial k^\beta}\ket{n}}{\epsilon_{mn}}+\frac{1}{\epsilon^2_{mn}}\left[{\sum_{c,\epsilon_c=\epsilon_n}}{v^\beta_{mc}O_{cn}}-{\sum_{c,\epsilon_c=\epsilon_m}}{ v^\beta_{cn}O_{mc}}\right]+\\+\frac{1}{\epsilon_{mn}}\left[{\sum_{c,\epsilon_c\neq\epsilon_m}}\frac{O_{mc}v^\beta_{cn} }{\epsilon_{mc}}-{\sum_{c,\epsilon_c\neq\epsilon_n}}\frac{ v^\beta_{mc}O_{cn}}{\epsilon_{cn}}\right].
\end{multline}

{
\subsection{Nonlinear responses for current}
In the above we showed a general formulation for treating second order responses. In this section, we demonstrate how our formulation naturally applies for the PT-even observables, in particular photocurrent. Using $O = e\hat{\mathbf{v}}$ with $\hbar{v}^\gamma_{nm}=\delta_{nm}\partial_\mathbf{k}^\gamma \epsilon_{n}+i\epsilon_{nm}A^\alpha_{nm}$, we find that Eqs.(\ref{eq:FullKI}) and (\ref{eq:FullKFS}) reduce to the standard resonant injection current \cite{PhysRevX.10.041041, PhysRevX.11.011001}:
\begin{multline}
    j^{\gamma}_{\rm injection}= \frac{\pi e^3}{4\eta}e^{2\eta t}\int_{\vec p}\int_\omega E^\alpha_{-\omega}E^\beta_{\omega}\times\\\times \sum_{nm}\delta(\hbar\omega-\epsilon_{mn})A^\beta_{mn}A^\alpha_{nm}f_{mn}(v^\gamma_{nn}-v^\gamma_{mm}), 
\label{injection}
\end{multline}
and injection current principal part \cite{PhysRevX.11.011001}:
\begin{multline}
    j^{\gamma}_{\rm injection~principal} = \frac{e^3}{2}\int_{\vec p}\int_\omega E^\alpha_{-\omega}E^\beta_{\omega}e^{2 \eta t}\times\\\times\sum_{nm}\mathcal{P}\frac{A^\beta_{mn}A^\alpha_{nm}f_{mn}}{(\hbar\omega-\epsilon_{mn})^2}(v^\gamma_{nn}-v^\gamma_{mm}).
\label{ferisea1}
\end{multline}
The set of shift nonlinearities as shown in Eqs.(\ref{shiftSI}) and (\ref{shiftFSSI}) is also consistent with Refs.\cite{PhysRevX.11.011001,PhysRevX.10.041041}. To see this for the shift photocurrent, we employ the identity:
\begin{multline}
    \partial^\beta_\mathbf{k} \frac{\hbar v^\gamma_{mn}}{\epsilon_{mn}}-i\left[\sum_{c,\epsilon_c\neq\epsilon_n}A^\beta_{mc}\frac{ \hbar v^\gamma_{cn}}{\epsilon_{cn}}-\sum_{c,\epsilon_c\neq\epsilon_m}\frac{ \hbar v^\gamma_{mc}}{\epsilon_{mc}}A^\beta_{cn}\right] =\\= i \partial^\gamma_\mathbf{k} A^\beta_{mn}+ (\sum_{c,\epsilon_c=\epsilon_m}A^\gamma_{mc}A^\beta_{cn}\hspace{-1mm}-\hspace{-1.75mm}\sum_{c,\epsilon_c=\epsilon_n}A^\beta_{mc}A^\gamma_{cn}),
\end{multline}
where we used $\partial^\gamma_\mathbf{k} A^\beta_{mn}\hspace{-0.75mm}-\hspace{-0.25mm}\partial^\beta_\mathbf{k} A^\gamma_{mn}\hspace{-1.5mm}=\hspace{-0.5mm}i\hspace{-0.5mm}\sum_c\hspace{-1mm}\left\{\hspace{-0.5mm}{A}^\gamma_{mc}A^\beta_{cn}\hspace{-1mm}-\hspace{-1mm}A^\beta_{mc}{A}^\gamma_{cn}\hspace{-0.5mm}\right\}$. Our formulation naturally takes care of degeneracies allowing to transparently produce the subtle U(2) gauge invariant form of circular shift currents in PT symmetric systems, see Ref.\cite{PhysRevX.11.011001}. Notice, for non-degenerate systems, condition $\epsilon_n = \epsilon_m$ implies $n=m$, thus we identify the shift vector as $ i \partial^\gamma_\mathbf{k} A^\beta_{mn}+ A^\beta_{mn}( A^\gamma_{mm}-A^\gamma_{nn})= A^\beta_{mn}R^{\gamma;\beta}_{mn}$ where $R^{\gamma;\beta}_{mn} = i\partial^\gamma_\mathbf{k}{\rm arg}[A^\beta_{mn}] + A^\gamma_{mm}-A^\gamma_{nn}$. Once substituted in Eq.(\ref{shiftSI}) we find the resonant shift photocurrent as in Eq.(4) of the Ref.\cite{PhysRevX.10.041041}. Notice that for $O = e\hat{\mathbf{v}}$, Eq.~(\ref{shiftFSSI}) produces an off-resonant principal part of the shift current; when combined with the injection current principal part, it produces a Fermi surface term consistent with Ref.\cite{PhysRevX.11.011001} and Ref.\cite{gao2021intrinsic}.}  

\section{Kramers quantum geometry and $\lambda$-formulation}\label{AsecD}
In this appendix we massage the KI, KFS and shift-like nonlinearities to their compact geometrical form as shown in the main text. As in the main text, we examine the set of parametric Hamiltonians $H_{\rm eff}(\lambda,\mathbf{p})= H_0(\mathbf{p}) + \lambda \hat O^{(-)}$, allowing us to write $\hat O^{(-)} = \partial H_{\rm eff}/\partial \lambda$, with the matrix elemets as in Eq.(\ref{eq:identity}) of the main text. 

It is important to note, that the $\lambda$-term lifts the exact PT symmetry protected degeneracy of Kramers states, since $\hat O^{(-)}$ is PT-odd. In order to explicitly track the states that in the limit of $\lambda\rightarrow0$ become Kramers degenerate, it is useful to denote such states via $\varepsilon_n\approx\varepsilon_c$. Naturally, in the limit $\lambda\rightarrow0$, the exact degeneracy of the equality is restored  $\approx \rightarrow =$. 
Using this notation we write the Kramers nonlinearities as:
\begin{multline}\label{lKr}
    \delta O^{(2)}_{\rm Kramers}= \frac{e^2}{2}\int_\omega E^\alpha_{-\omega}E^\beta_{\omega}\int_{\vec p}\sum_{nm}\Bigg[\frac{\mathcal{A}^\beta_{mn}f_{mn}}{(\hbar\omega-\varepsilon_{mn})^2+\eta^2}\times\\\times\left\{\sum_{c,\epsilon_c\approx\epsilon_n}\hspace{-2.25mm} O_{nc}\mathcal{A}^\alpha_{cm}\hspace{-0.75mm} -\hspace{-1.75mm} \sum_{c,\epsilon_c\approx\epsilon_m}\hspace{-2.25mm} \mathcal{A}^\alpha_{nc} O_{cm}\right\}\Bigg].
\end{multline}

Employing the covariant derivative with respect to $\lambda$ as:
\begin{equation}
S^{\alpha;\lambda}_{mn}\equiv\partial^{\lambda}S^\alpha_{mn}-i\Bigg[\hspace{-0.75mm}\sum_{c,\epsilon_c\approx\epsilon_m}\hspace{-2.25mm}{\mathcal{A}}^\lambda_{mc}S^\alpha_{cn}-\hspace{-0.75mm}\sum_{c,\epsilon_c\approx\epsilon_n}\hspace{-2.25mm}S^{\alpha}_{mc}{\mathcal{A}}^\lambda_{cn}\Bigg],
\end{equation}
we find the following identity:
\begin{multline}\label{covKr}
    \hbar\nabla_\lambda v^\alpha_{nm} - i\epsilon_{nm}\nabla_\lambda A^\alpha_{nm}=\\=i\sum_{c,\epsilon_c\approx\epsilon_n}O_{nc}\mathcal{A}^\alpha_{cm}-i\sum_{c,\epsilon_c\approx\epsilon_m}\mathcal{A}^\alpha_{nc} O_{cm}.
\end{multline}

In order to obtain the susceptibilities in the main text, we first note that the rectified responses of PT-odd observables to monochromatic electric fields of frequency $\Omega$ can be written generally as: 
\begin{multline}\label{helicitism}
    \delta O^{(2)} =
    e^2\underbrace{\left(\mathcal{E}^\alpha \mathcal{E}^{*\beta}+\mathcal{E}^{*\alpha} \mathcal{E}^{\beta}\right)}_{\text{helicity blind}}\frac12\left[\chi^{\alpha\beta}(\Omega)+\chi^{\alpha\beta}(-\Omega)\right]+\\+e^2\underbrace{\left(i\mathcal{E}^\alpha \mathcal{E}^{*\beta}-i\mathcal{E}^{*\alpha} \mathcal{E}^{\beta}\right)}_{\text{helicity sensitive}}\frac{1}{2i}\left[\chi^{\alpha\beta}(\Omega)-\chi^{\alpha\beta}(-\Omega)\right].
\end{multline}
After substitution of Eq.(\ref{covKr}) into Eq.(\ref{lKr}) and using Eq.(\ref{res_fs}) we find the KI and KFS susceptibilities displayed in Eq.(\ref{eq:genOing}) and  Eq.(\ref{eq:FS2}) of the main text respectively.  Notice that PT symmetry means that  $\chi_{\rm KI}(\omega) = -\chi_{\rm KI}(-\omega)$ and $\chi_{\rm KFS}(\omega) = -\chi_{\rm KFS}(-\omega)$ which combined with Eq.(\ref{helicitism}) implies that both are helicity dependent.

{We now turn to the shift resonant and shift Fermi sea response. Using the fact that $\sum_{c,\epsilon_c\neq \epsilon_n} = \sum_c - \sum_{c,\epsilon_c\approx \epsilon_n}$, and the following identity for the Berry connections: $\partial^\alpha_k \mathcal{A}^\lambda_{nm}-i\sum_c[{\mathcal{A}}^\alpha_{nc}\mathcal{A}^\lambda_{cm}-{A}^\lambda_{nc}\mathcal{A}^\alpha_{cm}]=\partial^\lambda \mathcal{A}^\alpha_{nm}$, we can rewrite two shift nonlinear susceptibilities as:
\begin{align}
    & \chi^{\alpha\beta}_{\rm SR}(\omega)\hspace{-0.75mm}=\hspace{-0.75mm} \frac{\pi}{2}\hspace{-0.75mm}\int_{\vec p}\sum_{nm}\hspace{-0.75mm}\delta(\hbar\omega\hspace{-0.75mm}-\hspace{-0.75mm}\varepsilon_{mn})f_{mn} i(\mathcal{A}^{\alpha;\lambda}_{nm}\mathcal{A}^\beta_{mn}\hspace{-1.25mm}-\hspace{-1mm}\mathcal{A}^\alpha_{nm}\mathcal{A}^{\beta;\lambda}_{mn})\label{eq:kramersshiftR}\\
    & \chi^{\alpha\beta}_{\rm SFS}(\omega)\hspace{-0.75mm}=\hspace{-0.75mm}  \frac{1}{2}\hspace{-0.75mm}\int_{\vec p}\sum_{nm} (\mathcal{A}^{\alpha;\lambda}_{nm}\mathcal{A}^\beta_{mn}\hspace{-0.75mm} + \hspace{-0.75mm}\mathcal{A}^\alpha_{nm}\mathcal{A}^{\beta;\lambda}_{mn}) \mathcal{P}\frac{f_{nm}}{\hbar\omega-\varepsilon_{mn}}.\label{eq:kramersshiftFS}
\end{align}
Much like Eq.~(\ref{eq:genOing}) of the main text, both shift resonant and shift fermi sea nonlinearities in Eqs.~(\ref{eq:kramersshiftR}) and (\ref{eq:kramersshiftFS}) depend on a generalized covariant derivative that includes the Kramers pairs. Importantly, PT symmetry restricts shift resonant response to be helicity blind: $\chi_{\rm SR}(\omega) = \chi_{\rm SR}(-\omega)$, and thus it identically vanishes in Fig.(\ref{FIG2}) of the main text and Fig.(\ref{FIGS1}). In contrast however, the shift Fermi sea is helicity sensitive, since for PT, we have $\chi_{\rm SFS}(\omega) = -\chi_{\rm SFS}(-\omega)$. Unlike Eq.~(\ref{eq:genOing}) of the main text, Eq.~(\ref{eq:kramersshiftFS}) can be non-vanishing for PT-symmetric non-degenerate energy bands} only leading to the restriction $\chi^{\alpha\beta}_{\rm SFS}(\omega)=\chi^{\beta\alpha}_{\rm SFS}(\omega)=\chi^{\alpha\beta}_{\rm SFS}(-\omega)$. Notice, by replacing the label for the Hamiltonian parametrization $\lambda\rightarrow \mathbf{k}$ one reproduces the standard Shift nonlinearity as in Ref.\cite{PhysRevX.10.041041} and Ref. \cite{PhysRevX.11.011001} .

Interestingly, although both the KFS contribution in Eq.~(\ref{eq:FullKFS}) and the SFS contribution in Eq.~(\ref{eq:kramersshiftFS}) are intrinsic, their ratio does not appear to carry a particular universal significance. In analogy with the non-degenerate case, the KFS term tracks the $\lambda$-derivative of the band energy, whereas the SFS term involves the covariant $\lambda$-derivative of the interband Berry connection.

\section{MBT Hamiltonian}\label{AsecE}
In our numerical calculations of nonlinearities of MBT, we adopted low energy effective Hamiltonian~\cite{SuYangXuMBT,PhysRevLett.124.126402}, with the normal and AFM componenets of the Hamiltonian in the spin-orbital basis $(\ket{p_{z,{\rm Bi}}^+,\uparrow},\ket{p_{z,{\rm Te}}^-,\downarrow},\ket{p_{z,{\rm Te}}^-,\uparrow},\ket{p_{z,{\rm Bi}}^+,\downarrow})^T$. Explicitly the normal components of the Hamiltonian and the interlayer couplings are:
\begin{gather}
     { \renewcommand{\arraystretch}{1} h_N(\mathbf{k})\hspace{-0.75mm} =\hspace{-0.75mm} \epsilon_0(\mathbf{k})\mathbb{I}_4\hspace{-0.75mm}+\hspace{-0.75mm}\left(\hspace{-1mm}\begin{array}{cccc}
         m(\mathbf{k})&i\alpha k_-&&  \\
         -i\alpha k_+& - m(\mathbf{k})&&\\
         &&-m(\mathbf{k})&i\alpha k_-\\
         &&-i\alpha k_+&m(\mathbf{k})
    \end{array}\hspace{-1mm}\right),}\\ { \renewcommand{\arraystretch}{1} t=\left(\begin{array}{cccc}
         t_1&0&-\lambda&0  \\
         0&t_2&0&\lambda \\
         \lambda&0&t_2&0 \\
         0&-\lambda&0&t_1
    \end{array}\right)}
\end{gather}
with $k_\pm = k_x \pm ik_y$, the dispersive part $\epsilon_0(\mathbf{k})=\gamma_0+\gamma k^2$, and massive part $m(\mathbf{k})=m_0+\beta_0k^2$. The set of parameters adopted were extracted from Ref.\cite{SuYangXuMBT} that was recently used to capture the effective electronic behavior of antiferromagnetic even MBT layers. For the convenience of the reader, the parameters are as follows: $\gamma = 17.0~{\rm eV\cdot\AA^2}, m_0 = -0.04~{\rm eV},\beta_0 = 9.40~{\rm eV\cdot\AA^2}, \alpha=3.20~{\rm eV\cdot\AA}, m_1=0.05~{\rm eV},m_2=0.09~{\rm eV},t_1=-0.0533~{\rm eV},t_2 = 0.0463~{\rm eV},\lambda=0.0577~{\rm eV}$.

As we have in our model, the layer Hamiltonian is:
\begin{equation}
    h_l = h_N(\mathbf{k}) -(-1)^l h_{\rm AFM}(\mathbf{k}).
\end{equation}
{
\section{Comparison of PT-odd observable nonlinearities for different $\tau$}\label{AsecF} 
Kramers injection response in Eq.(\ref{eq:genOing}) is highly sensitive to the scattering time $\tau$. In the main text we considered $\tau=0.25$ ps as an illustration. Here, in Fig.(\ref{FIGS1}), we display additional simulation results for $\tau = 1{~\rm ps},~50{\rm~fs},~25{\rm~fs},~1{\rm~fs}$ to cover the range of realistic scattering times \cite{ma2022photocurrent}. Importantly, for larger scattering times, the Kramers nonlinearities become systematically dominant, showing that they are the leading layer polarization nonlinearities in MnBi$_2$Te$_4$, which is the same behavior as discussed in the main text.}

\bibliography{Kramers_bib}

\end{document}